\documentclass{osa-article}

\journal{osajournal}

\usepackage[utf8]{inputenc}
\usepackage{braket}
\usepackage{dsfont}
\usepackage{graphicx}

\newcommand{\mc}[1]{\mathcal{#1}}
\newcommand{\HS}[1]{\mc{H}_{#1}}

\newcommand{\vNorm}[1]{\left| #1 \right|}
\newcommand{\dmS}[1]{\mc{D}\left( \HS{#1} \right)} 
\newcommand{\hermS}[1]{\mc{S}\left( \HS{#1} \right)} 
\newcommand{\tr}[1]{\mathrm{tr}\left({#1}\right)}
\newcommand{\R}{\mathds{R}}
\newcommand{\M}[1]{\pmb{#1}}

\usepackage{lineno}

\articletype{Research Article}

\begin{document}

\title{Error-mitigated photonic variational quantum eigensolver using a single-photon ququart}


\author{Donghwa Lee,\authormark{1, 2, $^*$} 
Jinil Lee,\authormark{1, 2, $^*$} 
Seongjin Hong, \authormark{1}
Hyang-Tag Lim, \authormark{1, 2}
Young-Wook Cho, \authormark{1, 3}
Sang-Wook Han, \authormark{1, 2}
Hyundong Shin, \authormark{4}
Junaid ur Rehman, \authormark{1, 4, $\dag$}
and 
Yong-Su Kim, \authormark{1, 2, $\ddag$}
}

\address{\authormark{1}Center for Quantum Information, Korea Institute of Science and Technology (KIST), Seoul, 02792, Republic of Korea\\
\authormark{2}Division of Nano \& Information Technology, KIST School, Korea University of Science and Technology, Seoul 02792, Republic of Korea\\
\authormark{3}Department of Physics, Yonsei University, Seoul 03722, Republic of Korea\\
\authormark{4}Department of Electronics and Information Convergence Engineering, Kyung Hee University, Yongin 17104, Republic of Korea\\
\authormark{$^*$}These authors contributed equally to this work.
}

\email{\authormark{$\dag$}junaid@khu.ac.kr} 
\email{\authormark{$\ddag$}yong-su.kim@kist.re.kr} 



\begin{abstract}
We report the experimental resource-efficient implementation of the variational quantum eigensolver (VQE) using four-dimensional photonic quantum states of single-photons. The four-dimensional quantum states are implemented by utilizing  polarization and path degrees of freedom of a single-photon. Our photonic VQE is equipped with the quantum error mitigation (QEM) protocol that efficiently reduces the effects of Pauli noise in the quantum processing unit. We apply our photonic VQE to estimate the ground state energy of He--H$^{+}$ cation. The simulation and experimental results demonstrate that our resource-efficient photonic VQE can accurately estimate the bond dissociation curve, even in the presence of large noise in the quantum processing unit. 
\end{abstract}

\section{Introduction}



Ongoing efforts to build a useful quantum computer are currently in the noisy intermediate-scale quantum (NISQ) era, characterised by hardware with tens of qubits, noise in the evolution, and no error correction~\cite{preskill2018quantum}. 
On the hardware side, several proposals based on different hardware architectures are being pursued and actively developed, e.g., superconducting, trapped ion, and photonic systems \cite{arute2019quantum,benhelm2008towards,zhong2020quantum}. 
At the same time, quantum-classical hybrid algorithms---called variational quantum algorithms (VQAs)---are emerging as promising candidates for near-term practical use of quantum processors. VQAs have applications in a wide variety of fields ranging from chemistry to physics and machine learning~\cite{peruzzo2014photonic,o2016scalable,kandala2017hardware,yung2014trappedion,shen2017quantum,kokail2019self,biamonte2017quantum,lubasch2020variational,liang2020variational,benedetti2020hardware}. VQAs operate by preparing a parameterized trial state on the quantum processor and evaluating a cost function of interest by measuring the state. Then, the cost function is iteratively optimized by varying the parameters of the trial state on a classical computer followed by its preparation and measurement on the quantum processor. One example of VQAs is variational quantum eigensolver (VQE) where the cost function of interest is the expectation value of a Hermitian operator $H$. By minimizing this cost function, one may obtain the lowest eigenvalue of $H$~\cite{peruzzo2014photonic}. The key advantage of VQE is the ability of solving exponentially large $H$ by the linearly increasing number of qubits in the quantum processor. Since the first demonstration of VQE has been reported using a photonic system~\cite{peruzzo2014photonic}, various achievements in VQE are reported e.g., demonstration on different platforms \cite{o2016scalable,kandala2017hardware,yung2014trappedion,shen2017quantum,kokail2019self} and  introduction of quantum error mitigation (QEM)~\cite{barron2020measurement,su2020error}. 


Photonic systems have played an essential role in quantum information processing~\cite{knill2001scheme, hong2020experimental, wang2019boson, barreiro2008beating}. One attractive feature of photonic platforms is the possibility of utilizing multiple degrees of freedom e.g., polarization, spatial and temporal mode, of individual photons to encode multiple qubits on a single-photon~\cite{yoo2018experimental, wang201818}. Thus, we can increase the dimension of the Hilbert space without increasing the number of photons in the quantum processor. Note also that, some two-qubit gate operations can be easily implemented by addressing qubits to multiple degrees of freedom of a single-photon~\cite{yoo2018experimental}.




In this paper, we implement VQE using single-photon four-dimensional quantum states or photonic ququarts. The photonic ququart is demonstrated with polarization and spatial degrees of freedom. The two-qubit Hamiltonian of He--H$^+$ is encoded with the single-photon ququart and its ground state energy is obtained via VQE. Our photonic VQE is also accompanied by the QEM scheme which remedy the errors introduced by Pauli noise on both degrees of freedom. The experimental efficacy of the QEM scheme is verified with the existence of depolarizing noise with different strengths. Our results demonstrate the usefulness of the high-dimensional photonic quantum states in implementing quantum-classical hybrid algorithms.


\section{Theory}

In this section, we introduce the VQE procedure and discuss possible sources of noise and errors both in the classical and quantum processing. We also discuss a QEM scheme that is capable of mitigating the effects of Pauli noise in the VQE.

\subsection{Notations}
Let \( \HS{d}\) be the \( d\)-dimensional Hilbert space. A \( d\)-dimensional pure state \( \ket{\psi} \) is a \( d \times 1\) column vector in \( \HS{d}\) with the normalization condition \( \vNorm{\ket{\psi}} = 1\), where \( \vNorm{\cdot}\) is the vector (\( \ell_2\) ) norm. 
We denote the set of density operators on \( \HS{d}\), i.e., \( d \times d\) positive semidefinite operators with unit trace, as \( \dmS{d}\). 
By a slight abuse of notation, we sometimes denote the density operator of a pure state \( \ket{\psi}\) by its label alone without the ket, i.e., \( \psi = \ket{\psi}\bra{\psi}\). 
The set of Hermitian operators, i.e, \( H = H^{\dagger}\) where \( \left( \cdot \right)^{\dagger}\) is the conjugate transpose, is denoted by \( \hermS{d}\). 
We omit the domains of these operators when it is clear from the context. 
The expectation value of an observable \( H \) with respect to a state \( \rho \) is denoted by \( \braket{H}_{\rho}\), which is defined as \( \braket{H}_{\rho} = \tr{H \rho}\). 
The expectation value of \( H \) with respect to a pure state \( \ket{\psi} \) is  \( \braket{H}_{\psi} = \braket{\psi | H | \psi}\). 
We denote the eigenvalues of \(H \in \hermS{d}\) in nondecreasing order by \( E_\ell\), \( \ell \in \left\{ 0, 1, \cdots, d-1\right\}\). Then, \(E_0\) is the smallest eigenvalue of \( H \). 
We use the convention of denoting the identity operator on \( \HS{2} \) by \( \sigma_0\), and Pauli \(X, Y\), and \(Z\) operators by \( \sigma_1, \sigma_2\), and \( \sigma_3\), respectively. 

\subsection{Variational Quantum Eigensolver}
Variational quantum algorithms (VQAs) have emerged in the recent times as one of the leading candidates for providing application-oriented quantum computational advantage. At the heart of VQAs is the calculus of variations, i.e., introduction of small perturbations in the system in an attempt to find the maximum or the minimum of some property of interest. 

In this work, we are interested in the variational quantum eigensolver (VQE). The VQE is a special case of VQAs to estimate the eigenvalues of a given Hermitian operator. The given Hermitian operator \( H \in \hermS{d}\) may correspond to the electronic structure of some molecule of interest, then the lowest eigenvalue of \( H\) corresponds to the ground state energy of this molecule. In particular, VQE attempts to estimate \( E_0 \) by minimizing the Rayleigh quotient 
\begin{align}
	R\left( H, \ket{\psi} \right) = \frac{\braket{\psi | H | \psi}}{\braket{\psi | \psi}},
\label{eq_H}
\end{align}
 by varying \( \psi \in \dmS{d}\) and utilizing the fact that \( E_0 \leq R\left( H, \ket{\psi} \right)\). Note that since we work with normalized states, \( \braket{\psi | \psi} = 1\), we can ignore the denominator. Then, the Rayleigh quotient reduces to the expectation \( \braket{\psi | H | \psi} = \braket{H}_{\psi}\).
 

The measurement of expectation \( \braket{H}_{\psi}\) is carried out by measuring a set of appropriate Pauli operators. It is known that any multi-qubit Hamiltonian can be decomposed into a number of Pauli operators with some weight coefficients, i.e.,
\begin{align}
	H = \sum_{j} w_j \M{\sigma}_j,
\end{align}
with \( w_j \in \R\), and \( \M{\sigma}_j\) are the Pauli strings, i.e., tensor products of multiple Pauli operators. Then, using the linearity of the expectation values, we can obtain
\begin{align}
	\braket{H}_{\psi} = \sum_j w_j \braket{\M{\sigma}_j}_{\psi}.
	\label{eq_H_exp}
\end{align}
That is, we can measure the expectation values of Pauli strings and then the appropriately weighted sum of these Pauli expectations gives an estimate on the expectation value of \(H\). 


\begin{figure}[t!]
	\centering
	\includegraphics[width=0.75\textwidth]{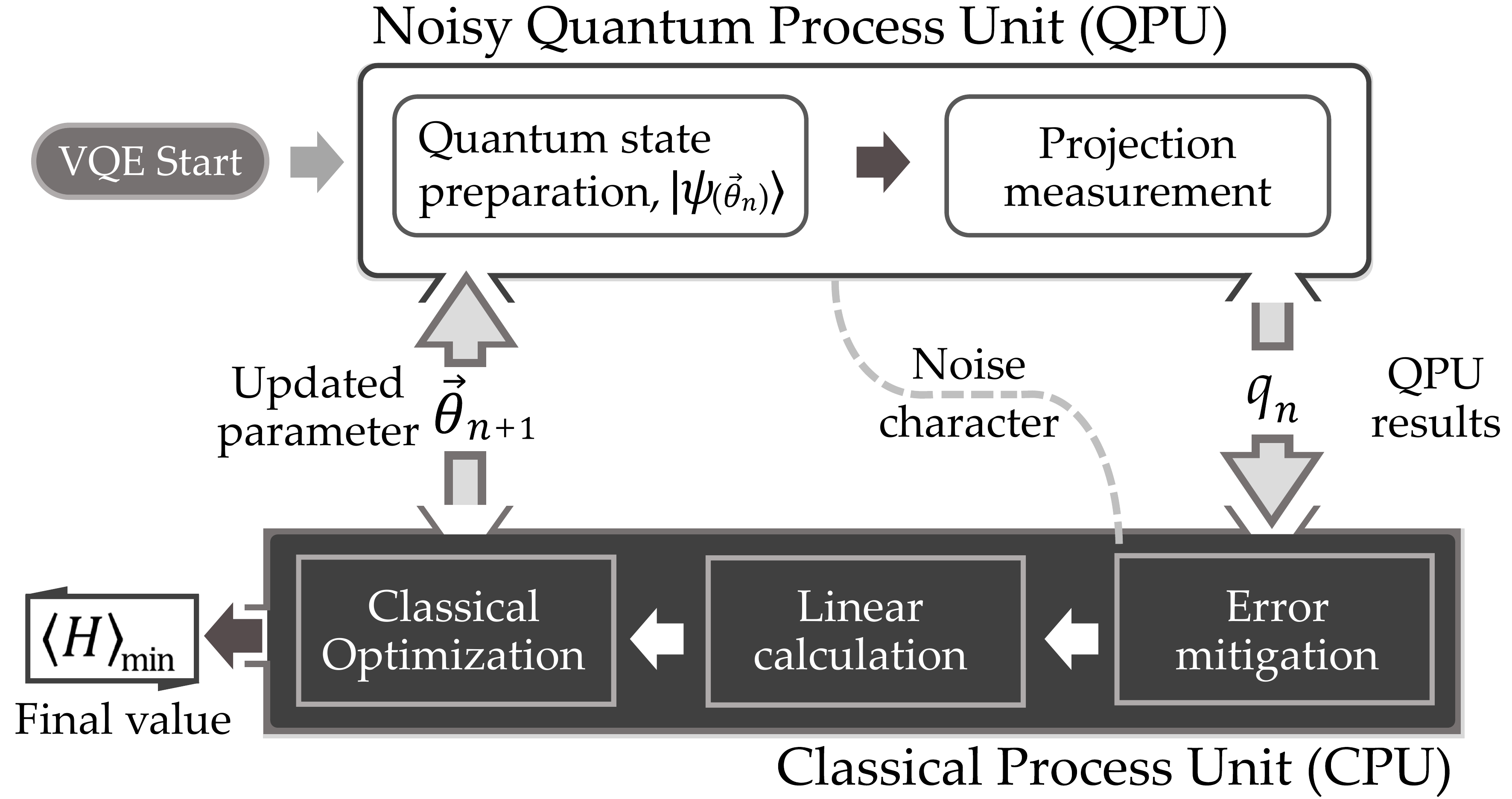}
	\caption{The VQE procedure. VQE estimates the minimum eigenvalue of an operator $H$ iteratively. Iteration $n$ includes preparing the state  $\vert\psi(\vec{\theta_{n}})\rangle$ and estimating the probabilities of measurement results on a quantum processing unit (QPU). This is followed by the classical steps that include quantum error mitigation (QEM), estimation of Pauli string expectations, and the estimation of $\langle H_n\rangle$. Finally, the estimated value of $\braket{H}_{\psi_n}$ is fed to an optimizer, which generates a new set of parameters $\vec{\theta}_{n+1}$. These steps are performed iteratively until the stopping criterion is satisfied. }
	\label{fig:fig1}
\end{figure}

Here, we use the VQE to estimate the ground state energy of $\mathrm{He}\text{--}\mathrm{H}^+$ cation \cite{peruzzo2014photonic}. The simplest form of Hamiltonian of this cation can be represented on two qubits, or on a single ququart as in our case. The Pauli decomposition of this Hamiltonian has nine terms, some of which can be measured simultaneously by nondegenerate projective measurements on \( \HS{4}\). Consequently, only four Pauli projective measurements are needed to estimate the expectation of our Hamiltonian of interest. See Supplement 1 for detailed Hamiltonian.

Finally, in order to minimize the expectation value, the target state is parameterized with a parameter set \( \vec{\M{\theta}}\in \R^{m}\), i.e., $\vert\psi\rangle = \vert\psi(\vec{\theta})\rangle$ . Ideally, \(m\) is linear, or at most polynomial in the number of qubits. At the start of the algorithm, the parameterized quantum state with random values assigned to these parameters is prepared. The expectation $\langle H \rangle _{\psi (\vec{\theta})}$ is estimated from the above procedure. The estimated value of this expectation is fed to a classical optimizer, which calculates a new set of parameters \( \vec{\theta} \). A new quantum state with these new parameters is prepared and a new estimate on the expectation is calculated with these parameters. This process is repeated several times until the estimated expectation value converges to a certain value. The minimum expectation value occurred during the above procedure is the estimate \(E_0\). Figure~\ref{fig:fig1} depicts the working procedure of a VQE including the QEM procedure which will be discussed later. Note that, an ideal VQE does not require QEM.

Ideally, VQE is able to accurately estimate the lowest eigenvalue \( E_0\) of the Hamiltonian of interest \( H\). On the other hand, practical implementations of the VQE may suffer from several non-idealities and noises. In the following, we discuss these sources of noise and the deviation of the practical VQE implementations from the ideal ones. 

\subsection{Noise and Errors in VQE}
In this section we outline different kinds of noise and errors in the implementation of VQE. In general, quantum error correction (QEC) is the standard approach for correcting errors in the implementation of quantum algorithms and protocols~\cite{NC:11:QCQI}. However, since the VQAs operate in the paradigm of NISQ computing, we may not have enough resources to implement full QEC. Furthermore, due to different nature of VQAs from general quantum algorithms, the encountered errors are also of different nature. 
We discuss these errors and approaches to remedy these errors in the following.

\subsubsection{Statistical Noise}
The estimation of Pauli expectations is achieved by preparing \( \ket{\psi}\) and measuring it in the eigenbasis of the corresponding Pauli string \( M \) times. Let \( s_j = \braket{\M{\sigma}_j}_{\psi}\) and \( \hat{s}_j\) be its estimate. Then, the difference \( \epsilon_j = s_j - \hat{s}_j\) is the error in the expectation estimation. Since \( M \) is finite, statistical noise will contribute to this error. Barring other sources of errors, it is intuitive to think that this error can be reduced by choosing a large \( M\). We make this notion more precise in the following by applying Hoeffding's inequality \cite{Hoe:63:JASA}. 

Since Pauli strings \( \M{\sigma}_j \) are products of Pauli operators, their spectrum is degenerate, i.e., it has only two eigenvalues -1 and +1. Then, without a loss of generality, we can assume \( M\) measurement outcomes \( X_1, X_2, \cdots, X_M\) to be independent and identically distributed Bernoulli random variable taking values from \( \left\{ -1, 1\right\}\). Then, 
\begin{align}
	\hat{s}_j = \frac{X_1 + X_2 + \cdots + X_M}{M},
\end{align}
and we can apply the Hoeffding's inequality to precisely bound the probability that the error \( \epsilon_j \) is greater than some positive number \( t \) as
\begin{align}
	P \left[ \left| \hat{s}_j - s_j\right| \geq t \right] \leq 2 \exp \left( -\frac{M t^2}{2}\right).
\end{align}
That is, most of the estimated results are concentrated around the true expectation value and the probability of deviating by \( t\) is exponentially small in \( M t^2\). Therefore, increasing the number of measurements \( M\) indeed exponentially decreases the probability of error being larger than some number \( t\).

Since the expectation \( \braket{H}\) is the weighted sum of the expectation of the Pauli strings, we can deduce that its estimates are also going to be concentrated around its true value. However, obtaining a precise bound on its deviation is more involved due to dependence of different Pauli expectations on each other and is beyond the scope of this paper. We only remark that increasing the number of measurements for expectation of each Pauli string reduces the statistical errors and improves the estimates of \( \braket{H}\).

In our experiments we use \( M \approx 4,000\) measurements per Pauli string. This leaves us with the probability less than 0.013 that the error in estimating \( \braket{\M{\sigma}_j}_{\psi}\) is greater than 0.05.

\subsubsection{Optimization Errors}
Another source of possible errors in a VQE implementation can be the optimization errors. We define the optimization errors to be the errors where the classical optimizer returns a minimum of the objective function which is not the true minimum. These errors can be further divided into two separate categories i) inherent problems of the optimizer, and ii) optimization errors due to the presence of local minima or vanishing gradients in the landscape of the objective function. We discuss both of these categories in some detail in the following. 

In the general implementations of the VQAs, the cost function is not analytically obtained, which makes it difficult to utilize some gradient-based optimizers. There are a few studies to utilize the gradient-based optimizers for VQAs with some success~\cite{SWM:20:Qua, SIK:20:Qua}. However, these methods suffer from the vanishing gradients and generally perform worse than the gradient-free optimization methods due to noise~\cite{UB:20:arx, WFC:21:arx}. Therefore, VQAs often employ heuristics-based optimizers including Nelder-Mead, Powell, COBYLA, and L-BFGS-B. A comparison of their performance for solving the ground-state energy with VQE can be found in Ref.~\cite{RBM:18:arx, PSP:21:QIP}.
	
One problem with these heuristics-based optimizers is that they cannot guarantee the closeness of the obtained solution from the actual solution. Then, one has to carefully choose the stopping criterion for the optimizer. This stopping criterion can be in the form of total number of quantum processing unit (QPU) calls, some convergence behavior of the objective function values, or a combination of both. 


One can only hope to obtain a good quality solution by allowing a large number of QPU calls. However, it is not possible to exactly characterize the number of QPU calls in terms of required quality of solution. Then, the only possibility is to find the sufficient number of QPU calls by hit-and-trial. For our case, we found $50\sim200$ QPU calls to be sufficient to obtain the minimum of our objective function depending on the optimizer, see Supplement 2 for the test results of various optimizers.

The second challenge is the vanishing gradients of the objective function in the space of state parameters. This phenomenon is referred to as barren plateus in the landscape of objective function. It can be caused by the nature of the objective function as well as due to the noise in the quantum hardware~\cite{UB:20:arx, WFC:21:arx}. It is being actively studied and new techniques are being developed to understand and reduce its effects in optimizing parameterized quantum circuits. Some possible remedies to avoid the barren plateus include by appropriately modifying the cost function, detecting the existence of barren plateus using the quantum control theory, and trading off the size of the solution space in favor of convergence by reducing the expressibility of ansatz states~\cite{CSV:21:npjQI, LCS:21:arx}.

\subsubsection{Quantum Noise}
Quantum noise is the noise present in the state preparation, evolution, and measurement, i.e, anywhere in the QPU. Traditionally, readout noise is treated differently from the noise in state preparation and state evolution. However, as we show below, the noise present in our system can be `absorbed' in the readout errors. Consequently, we are able to employ a unified error mitigation scheme for mitigating all types of noise in our QPU.

A quantum channel \( \mc{N}\), i.e., a trace-preserving completely positive (TPCP) map, can be represented in terms of its Kraus operators with an input state \( \rho \) as
\begin{align}
	\mc{N} \left( \rho \right) = \sum_j K_j \rho K_j^{\dagger},
\end{align}
where \( K_j\) are the Kraus operators satisfying \( \sum_j K_j^{\dagger}K_j=I\) where $I$ is the identity matrix. We use photonic four-dimensional states in our experiments, which consists of the single-photon polarization and path degrees of freedom. The most common quantum noise of this system is depolarization and dephasing noise. We can model these type of noise as two (possibly correlated) Pauli channels on each degree of freedom, i.e., 
\begin{align}
	\mc{N}_{a}\otimes\mc{N}_{b}\left( \rho_{a, b}\right) = \sum_{j, k} p_{j, k} K_j^a \otimes K_k^b \rho_{a, b} \left( K_j^a \otimes K_k^b\right)^{\dagger},
\end{align}
where the subscripts a and b represent polarization and path degrees of freedom, respectively. Furthermore, 
\begin{align}
	K_0^{a} &= K_0^{b} =  \begin{bmatrix}
		1 & 0\\ 0 & 1
	\end{bmatrix}, \ 
	K_1^{a} = K_1^{b} =  \begin{bmatrix}
		0 & 1\\ 1 & 0
	\end{bmatrix}, \nonumber \\
	K_2^{a} &= K_2^{b} =  \begin{bmatrix}
		0 & -\dot{\iota}\\ \dot{\iota} & 0
	\end{bmatrix}, \,
	K_3^{a} = K_3^{b} =  \begin{bmatrix}
		1 & 0\\ 0 & -1
	\end{bmatrix},
\end{align}
where \( \dot{\iota} = \sqrt{-1}\) and \(  p_{j, k}\) is the probability that the operator \(  K_j^a \otimes K_k^b\) acts on the input state. If the two channels are uncorrelated, we have \( p_{j, k} = q_j^{a}q_k^{b}\). 


Readout errors in the measurement of a quantum state are modeled as a left stochastic matrix~\cite{MZO:20:qua}. By letting \( p\) and \( q\) be the ideal and measured (noisy) probabilities obtained after measuring some state \( \rho\), we have,\begin{align}
	q = \Lambda p,
	\label{eq:meas_noise}
\end{align}
where \(  \Lambda\) is the left stochastic matrix, i.e., \( \Lambda_{j, k} \in \left[ 0, 1\right]\), \( \sum_j \Lambda_{j, k} = 1\), which models the behavior of a noisy measurement device~\cite{MZO:20:qua}. 
To see the effect of Pauli noise on Pauli measurements, consider the following scenario \cite{rehman2021entanglementfree}. Let us assume that we are interested in measuring the Pauli string \( P\) on a quantum state \( \rho\). We can decompose \( \rho\) in the eigenbasis of \( P\) as
\begin{align}
	\rho &= \sum_{j, k} \alpha_{j, k}\ket{\phi_{j}}\bra{\phi_k}\nonumber \\
	&= \sum_{j}\alpha_{j, j}\ket{\phi_{j}}\bra{\phi_j} + \sum_{\substack{j, k\\ j \neq k}} \alpha_{j, k}\ket{\phi_{j}}\bra{\phi_k},
	\label{QEM1}
\end{align}
where \( \left\{ \ket{\phi_j}\right\}_j\) is the eigenbasis of \( P\), and \( \alpha_{j, j} \in \R\). Clearly, the off-diagonal terms on the right side of Eq.~\eqref{QEM1} have no contribution in the measurement outcomes and the probability of obtaining measurement outcome \( j\) in this setup is \( p_j \coloneqq \alpha_{j, j}\). We define \( p^{\mathrm{noiseless}} = [p_j]_j\).


Since \(n\)-qubit Pauli channel is a random unitary channel, its effect on \( \rho \) is to randomly apply one of the Pauli string on \( \rho\). Furthermore, there does not exist any Pauli string \( P_j\) such that \( P_j \ket{\phi_i}\bra{\phi_k}P_j^{\dagger} = \ket{\phi_x}\bra{\phi_x} \) for any \(j, i \neq k \), and \( x\). Therefore, the off-diagonal terms remain irrelevant even after passing through a Pauli channel, and we can assume \( \rho \approx \tilde{\rho} = \sum_j \alpha_{j, j} \ket{\phi_{j}}\bra{\phi_j}\).

It was shown in Ref.~\cite{RJK:18:SR, RJS:19:PRA} that the effect of a discrete Weyl channel (a generalization of Pauli qubit channels) on the eigenstates of a Weyl operator can be modeled as a classical symmetric channel.\footnote{These results were later generalized for generalized Pauli channels \cite{SIU:20:JPAMT}.}  Since the state of interest \( \tilde{\rho}\) is diagonal in the eigenbasis of \( P\), the same argument of modeling the Pauli channel as a classical symmetric channel holds here. The transition matrix of a classical symmetric channel is doubly stochastic. Then, the vector of measurement probabilities with an ideal detector after the effect of Pauli channel is
\begin{align}
	p^{\mathrm{noisy}} = \Delta p^{\mathrm{noiseless}},
\end{align}
where \( \Delta\) is the doubly stochastic matrix representing the effect of Pauli channel. Finally, the effects of measurement errors on \( p^{\mathrm{noisy}}\) can be accounted for by substituting \( p \) in Eq.~\eqref{eq:meas_noise} with the last equation, i.e.,
\begin{align}
	q = \Lambda \Delta p^{\mathrm{noiseless}} = \Gamma p^{\mathrm{noiseless}},
	\label{eq:meas_Pauli_noise}
\end{align}
where we have defined \( \Gamma = \Lambda \Delta\). Finally, we recall that a doubly stochastic matrix is also left stochastic and the product of two left stochastic matrices is again a left stochastic matrix. Hence, \( \Gamma\) is also a left stochastic matrix and Eq.~\eqref{eq:meas_Pauli_noise} has the exact same form as that of a noisy detector characterized by \( \Gamma\) with a noiseless state. 

This formulation allows us to treat the errors from the noisy evolution as well as from the measurement noise in a unified manner as measurement errors alone. Consequently, we only need to perform quantum detector tomography, which is equivalent to Pauli channel tomography in this formulation, to obtain \( \Gamma_j\) for each Pauli string \( P_j\) that we want to measure. Then, we can perform QEM based on these \( \Gamma_j\), which mitigate the errors arising from noisy evolution as well as from the measurement noise \cite{rehman2021entanglementfree}. 



Now let us discuss how one can experimentally construct \( \Gamma_j \). We begin to prepare eigenstates of \( P_j\). These eigenstates are allowed to evolve naturally under the system noise, and then the measurement in the eigenbasis of \( P_j \) is performed. Then, the \( (k, \ell) \)-th entry of \( \Gamma_j \) is the relative frequency of measuring the \( \ell\)-th eigenstate of \( P_j\), when \(k\)-th eigenstate of \( P_j\) was prepared. Once all \( \Gamma_j\) corresponding to the \( P_j\) whose measurement is required for Hamiltonian estimation are obtained, we can perform the QEM by the following two methods \cite{MZO:20:qua}. We directly invert  \( \Gamma_j\) to obtain the vector of error mitigated probabilities \( p_j^{\mathrm{mit}}\) from the experimentally obtained vector of probabilities \( p_j^{\mathrm{exp}}\), i.e., \( p_j^{\mathrm{mit}} = \Gamma_j^{-1} p_j^{\mathrm{exp}}\). However, this may result into a vector which is not a correct probability vector, i.e., elements may negative or the sum may be greater 1. In such cases, we can still obtain a valid vector of probabilities by projecting the obtained vector back onto the probability simplex.

We remark that this formulation and mitigation of Pauli channel errors in the framework of measurement errors was recently proposed in \cite{rehman2021entanglementfree} and our work also serves as the first experimental validation of this approach. {Moreover, while the original proposal was proposed in the context of parameter estimation of generalized Pauli channels, here, we have shown that this QEM scheme is also applicable to the VQE and other NISQ algorithms where only Pauli measurements are utilized.} This formulation requires no additional qubits or quantum resources, except for the estimation of the noise matrices \( \{\Gamma_j\}_j \). Furthermore, this method of QEM produces reasonable estimates of the objective function as long as the noise matrices \( \Gamma_j \) are nonsingular. This condition is satisfied, e.g., when the noise is not maximally depolarizing or completely dephasing.


\section{Experiment}

\begin{figure}[t]
\includegraphics[angle=0,width=5in]{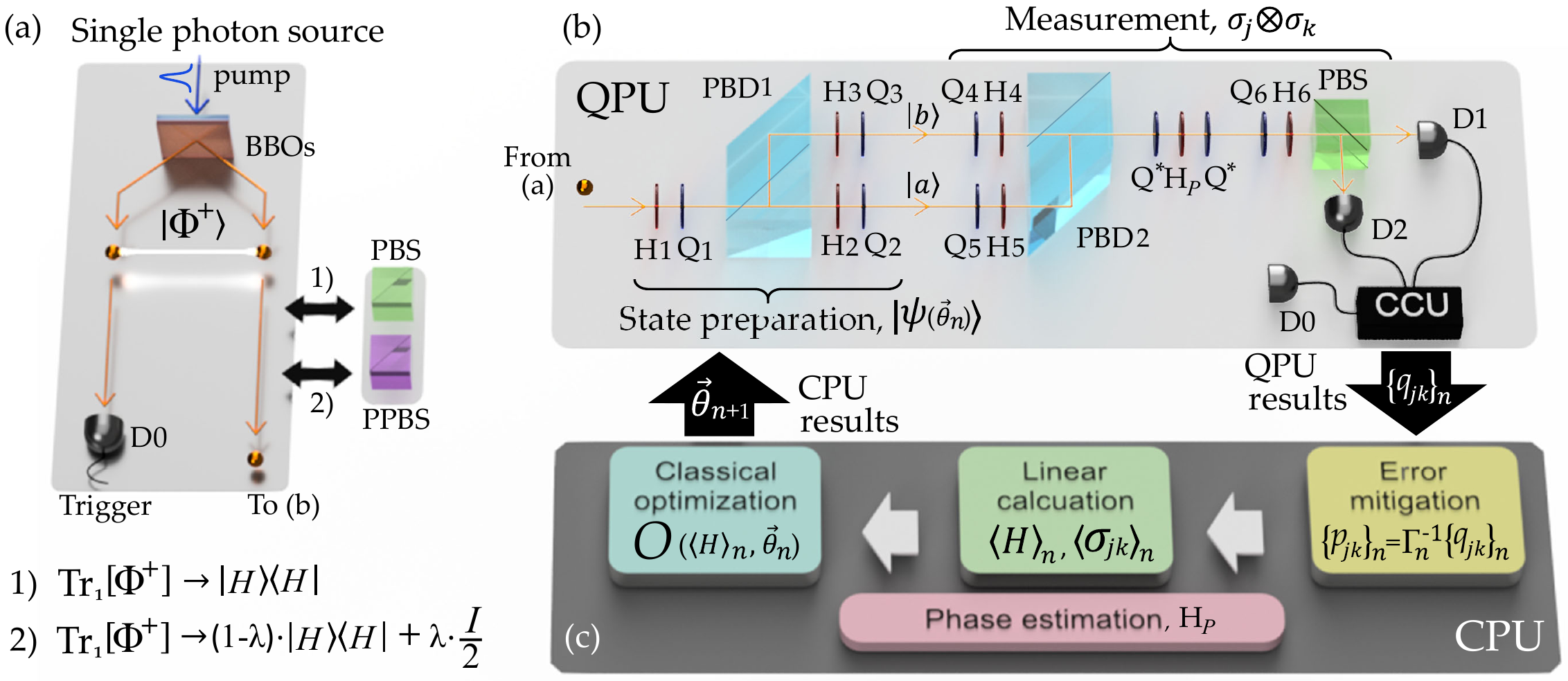}
\caption{The experimental setups of (a) single-photon generation, (b) QPU using photonic ququarts, and (c) classical processing unit (CPU). A two-photon Bell state is generated by SPDC and one photon is detected to herald a singe photon which is sent to the VQE setup. The input polarization state is filtered either by a PBS or PPBS. The polarization filtering with PPBS causes the depolarizing noise in polarization degree of freedom. H: half waveplate, Q: quarter waveplate, PBD: polarization beam displacer, PBS: polarizing beamsplitter, D: avalanche photodioide, CCU: coincidence counting unit.}
\label{vqe_setup}
\end{figure}

Figure~\ref{vqe_setup} presents the overall experimental setup of our photonic VQE using four-dimensional quantum states. The heralded single-photon state is prepared via spontaneous parametric down-conversion and a single-photon detection at the trigger detector D0, see Fig.~\ref{vqe_setup}(a). We have initially generated a Bell state $\vert\Phi^{+}\rangle=\frac{1}{\sqrt{2}}(\vert HH \rangle + \vert VV \rangle)$ using a sandwich BBO crystals where $\vert H \rangle$ and $\vert V \rangle$ denote the horizontal and vertical polarization states, respectively~\cite{wang2016experimental,pramanik2019revealing,pramanik2019nonlocal}. Then, the polarization state of the heralded single-photon is prepared to $|H\rangle$ using a polarizing beamsplitter (PBS), and it is sent to the photonic QPU shown in Fig.~\ref{vqe_setup}(b).




The incoming single-photon state to the QPU is split into two spatial modes $|a\rangle$ and $|b\rangle$ using sets of half- and quarter-waveplates (HWP, QWP) noted as H1 and Q1, and a polarizing beam displacer (PBD1) which transmits and reflects $|H\rangle$ and $|V\rangle$, repectively. The amplitude and relative phase of the spatial modes $|a\rangle$ and $|b\rangle$ can be tuned with H1 and Q1. Then, sets of waveplates, H2, Q2, H3, and Q3, placing at each spatial mode encode the polarization state at each spatial mode, and thus, we can generate arbitrary four-dimensional quantum states of
\begin{eqnarray}
\vert \psi(\vec{\theta})\rangle = \alpha\vert aH \rangle+\beta\vert aV \rangle+\gamma\vert bH \rangle+\delta\vert bV \rangle,
\label{4-d state}
\end{eqnarray}
where $|\alpha|^2+|\beta|^2+|\gamma|^2+|\delta|^2=1$. Considering the normalization condition and global phase in Eq.~(\ref{4-d state}), the six waveplates $\{{\rm H1,Q1,H2,Q2,H3,Q3}\}$ span all four-dimensional pure quantum states~\cite{yoo2018experimental}.

The projection measurements onto Pauil strings, $\sigma_{j}\otimes\sigma_{k}$, can be performed by sets of waveplates $\lbrace${\rm H4, Q4, H5, Q5, H6, Q6}$\rbrace$ and PBD2. In order to compensate the phase drift between the spatial modes $|a\rangle$ and $|b\rangle$, we have inserted a HWP, $\rm{H_p}$, between two QWPs at $45^\circ$ after the PBD2~\cite{yoon2019experimental}. During the experiment, we have measured and compensated the phase drift every 10 minutes. Finally, single photons are detected by avalanche photodiodes D1 and D2 and two-fold coincidences between the trigger detector D0 and D1 or D2, D01 and D02 are registered using a home-made coincidence counting unit (CCU)~\cite{park2015high, park2021arbitrary}. The optical setup of the VQE experiment is tested with the ability to generate and measure arbitrary four-dimensional quantum states with high purity, $P>0.98$. Examples of generated four-dimensional quantum states are given in Supplement 3.


The classical processing unit (CPU) receives the measurement outcomes from QPU and performs QEM, linear calculation, and classical optimization. The QEM procedure proposed in theory can be optionally applied before the linear calculation. The CPU calculates the expectation value of Hamiltonian $\langle H\rangle_n$ according to Eq.~(\ref{eq_H_exp}). Then, with the input of $\{\vec{\theta}_n,\langle H\rangle_n\}$, a classical optimizer updates the input parameters, $\vec{\theta}_{n+1}$, to find the minimum expectation value $\langle H\rangle$.  In our experiment, we alter six angles of waveplates, $\vec{\theta}_n=\{{\rm H1, Q1, H2, Q2, H3, Q3}\}$ on the state preparation. 

\begin{table}[t]
\setlength{\tabcolsep}{0.15in}
\centering
\begin{tabular}{|c|c|c|}
\hline
Classical optimizer & $P_S$ & $\overline{N}$\\
\hline
Nelder-Mead & 0.52  & 130 \\
Powell & 0.94  & 112 \\
COBYLA & 0.97  & 51 \\
\hline
\end{tabular}
\caption{The summarized emulation results of classical optimizers. It was tested with the Hamiltonian $H$ at the interatomic distance $R=0.9~$\AA~ and the probability was obtained from 1,000 independent trials. $P_S$ and $\overline{N}$ denote the success probability to find the minimum eigenvalue with the error of less than 0.01~MJ/mol, and average number of iterations, respectively. It shows that COBYLA presents the best performance in terms of both success probability and number of iterations.}
\label{optimizer}
\end{table}
The performance of VQE is highly dependent on the classical optimizer. In our experiment, we have tested three simplex-based direct search methods, Nelder-Mead, Powell, and COBYLA. We have emulated the performance of these optimizers using a classical computer, and summarized the results in Table~\ref{optimizer}. It is obvious that COBYLA provides the best performance in terms of both success probability $P_S$ and number of iterations $\overline{N}$. We have also verified the performance of different optimizers with real VQE experiments. Therefore, in the following VQE experiment, we have utilized COBYLA for a classical optimizer. The details of the emulation and experimental test results for the classical optimizers are presented in Supplement 2.

\begin{figure*}[b]
\includegraphics[angle=0,width=4.5in]{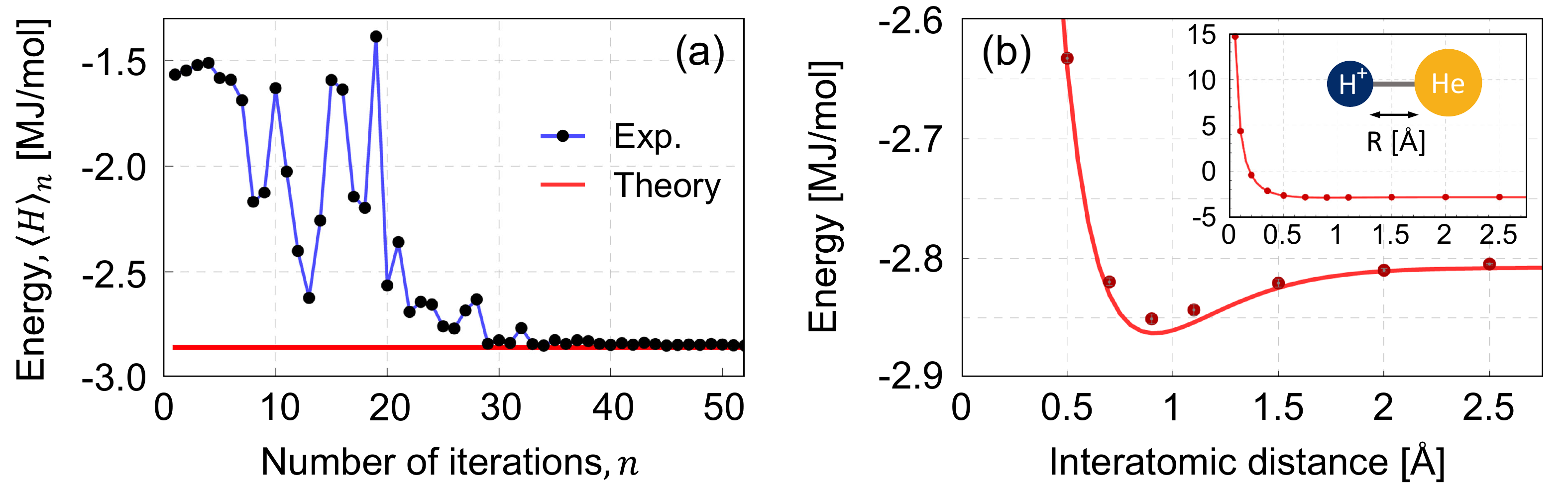}
\centering
\caption{The experimental results of VQE using photonic ququart state. The red line and circles indicate the theory value and estimated bound energy via VQE, respectively. (a) The estimated bound energy at $R=0.9~$\AA~ changes in the single VQE run with respect to the number of iterations $n$.  The estimated bound energy converges to the theory value as the number of iteration increases. (b) The estimated bound energy with respect to the interatomic distance.} 
\label{vqe_graph}
\end{figure*}

\begin{figure*}[t]
\includegraphics[angle=0,width=4.5in]{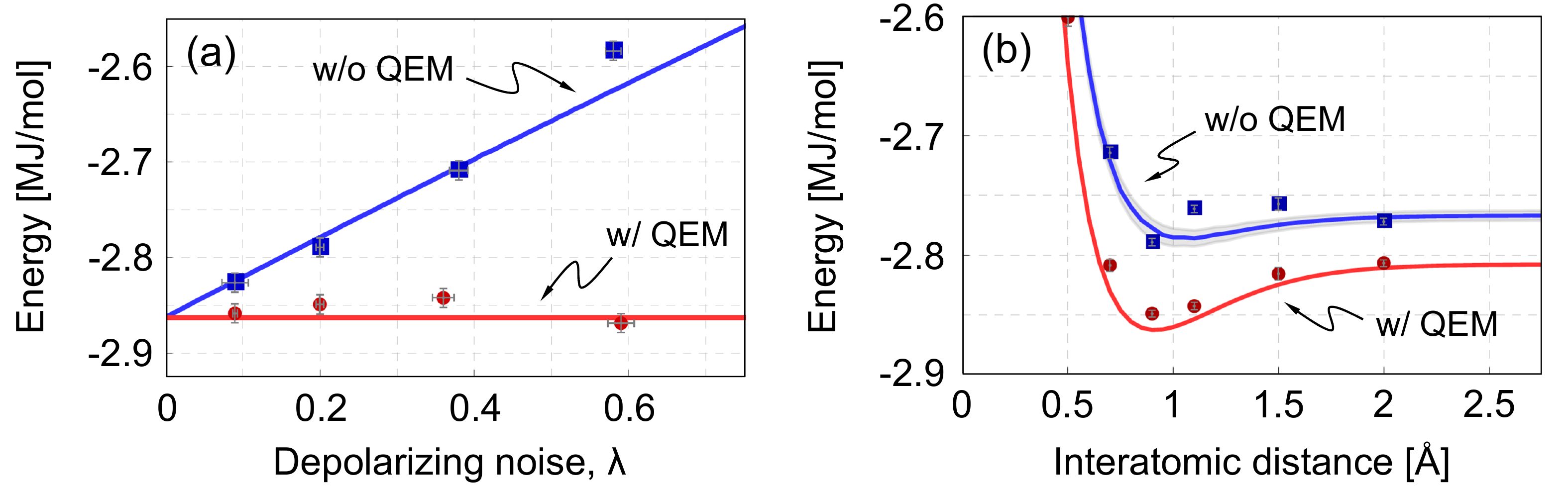}
\centering
\caption{The VQE experimental results with depolarizing noise. Red lines present the theoretical bound energy with ideal QPU. Blue lines denote the expected bound energy with corresponding amount of depolarizing noise $\lambda$. Red circles and blue squares are the experimental data with and without QEM protocol. The errors in the amount of depolarizing noise presents the experimental standard deviation from Pauli channel tomography. (a) The estimated bound energy with respect to the amount of depolarizing noise $\lambda$. While the depolarization noise introduces discrepancy between theoretical and experimental bound energies, one can successfully find the correct bound energy with our QEM protocol. (b) The estimated bound energy as a function of interatomic distance with depolarizing noise of $\lambda=0.20\pm0.02$.}
\label{vqe_mit_graph}
\end{figure*}


In order to verify the performance of our photonic VQE using a single-photon ququart, we run VQE to estimate the ground energy of He--H$^{+}$ cation without QEM protocol. Figure~\ref{vqe_graph} (a) shows the estimated energy expectation $\langle H\rangle_n$ at the interatomic distance $R=0.9~$\AA~ with respect to the number of iterations $n$. During the iteration, the estimated energy expectation converges to the theoretical value presented as a red straight line. Figure~\ref{vqe_graph} (b) shows the estimated bound energy of He--H$^{+}$ with respect to the interatomic distance $R$. The red line and circles are the theoretical and the experimental values for given interatomic distances. The experimental values and error bars are obtained by taking averages and standard deviations of the five minimum points during a single VQE run. We note that the error bars size is smaller than the markers in Fig.~\ref{vqe_graph} (b). It clearly shows that the bound energy estimated by our VQE experiment is well agreed with the theory. The experimentally obtained minimum ground state energy, $E_{g}~=~$-2.848 $\pm$ 0.004 MJ/mol at $R~=~0.9~$\AA, is close to its theoretical value of $E_{\rm th}=-2.863$ MJ/mol. 

In order to verify the effectiveness of our QEM protocol, we prepared initial states with the depolarization noise as $\rho=(1-\lambda)\cdot\vert H \rangle\langle H \vert +\lambda\cdot\frac{I}{2}$, where $I$ is the $2\times2$ identity matrix. The noisy states is prepared by replacing the PBS at the single photon source setup to partial PBS (PPBS) which partly transmits vertical polarization state~\cite{pramanik2019revealing}. The quantum channel was characterized by the Pauli channel tomography. With the results, we construct the error mitigation matrix $\Gamma_n$ on each Pauli basis and utilized it for QEM.


Figure~\ref{vqe_mit_graph} (a) presents the calculated bound energy at $R=0.9~$\AA~ with respect to the amount of the depolarizing noise $\lambda$. Without QEM protocol, the calculated bound energy increases as the noise increases. On the other hand, our QEM protocol successfully finds correct results regardless of the amount of noise. Note that, as discussed in theory, our QEM protocol can efficiently remedy noise except for $\lambda=1$. Figure~\ref{vqe_mit_graph} (b) shows the bound energy with respect to the interatomic distance $R$ with the noise of $\lambda=0.20\pm0.02$. These experimental results clearly show the effectiveness of our QEM protocol which successfully finds the correct results without further quantum resources.



\section{Conclusions}

We have presented an experimental implementation of variational quantum eigensolver on a photonic ququart to estimate the ground state energies of He--H$^+$ cations. Using polarization and path degrees of freedom of a single-photon, we were able to encode two-qubit Hamiltonians on a single-photon. Additionally, we employed the Pauli channel estimation and QEM scheme to reduce the effects of noise in QPU and obtain more accurate estimates of ground state energies. Our results clearly show that high-dimensional photonic quantum states based on multiple degrees of freedom provide a resource-efficient way to implement variational quantum algorithms. 

We remark that the experimental demands of some entangling operations in multiple degrees of freedom encoding system become comparable to that of single-qubit operation. For instance, in our photonic system, the experimental difficulty of Bell state measurement, which projects the input quantum states to four Bell states, becomes comparable to simple Pauli string projection measurement. This feature suggests our system as a natural choice to recently proposed scheme to reduce the number of measurement settings via entangling measurements~\cite{hamamura2020efficient}. We also propose some possible future research directions including utilization of more degrees of freedom to make this implementation even more efficient and expanding the QEM scheme to more general quantum noise models. 


\section*{Funding}

National Research Foundation of Korea (2019M3E4A1079777, 2019R1A2C2006381, 2019R1I1A1A01059964, 2021R1C1C1003625, and 2019R1A2C2007037, 2021M1A2A2043892); MSIT/IITP (2020-0-00972 and 2020-0-00947); KIST research program (2E31021).

\section*{Disclosures}
The authors declare no conflicts of interest.


\begin{thebibliography}{10}
\newcommand{\enquote}[1]{``#1''}

\bibitem{preskill2018quantum}
J.~Preskill, \enquote{Quantum computing in the {NISQ} era and beyond,}
  {\protect\JournalTitle{Quantum}} \textbf{2}, 79 (2018).

\bibitem{arute2019quantum}
F.~Arute, K.~Arya, R.~Babbush, D.~Bacon, J.~C. Bardin, R.~Barends, R.~Biswas,
  S.~Boixo, F.~G. Brandao, D.~A. Buell \emph{et~al.}, \enquote{Quantum
  supremacy using a programmable superconducting processor,}
  {\protect\JournalTitle{Nature}} \textbf{574}, 505--510 (2019).

\bibitem{benhelm2008towards}
J.~Benhelm, G.~Kirchmair, C.~F. Roos, and R.~Blatt, \enquote{Towards
  fault-tolerant quantum computing with trapped ions,}
  {\protect\JournalTitle{Nature Physics}} \textbf{4}, 463--466 (2008).

\bibitem{zhong2020quantum}
H.-S. Zhong, H.~Wang, Y.-H. Deng, M.-C. Chen, L.-C. Peng, Y.-H. Luo, J.~Qin,
  D.~Wu, X.~Ding, Y.~Hu \emph{et~al.}, \enquote{Quantum computational advantage
  using photons,} {\protect\JournalTitle{Science}} \textbf{370}, 1460--1463
  (2020).

\bibitem{peruzzo2014photonic}
A.~Peruzzo, J.~McClean, P.~Shadbolt, M.-H. Yung, X.-Q. Zhou, P.~J. Love,
  A.~Aspuru-Guzik, and J.~L. O'brien, \enquote{A variational eigenvalue
  solver on a photonic quantum processor,} {\protect\JournalTitle{Nature
  Communications}} \textbf{5}, 1--7 (2014).

\bibitem{o2016scalable}
P.~J. O'Malley, R.~Babbush, I.~D. Kivlichan, J.~Romero, J.~R. McClean,
  R.~Barends, J.~Kelly, P.~Roushan, A.~Tranter, N.~Ding \emph{et~al.},
  \enquote{Scalable quantum simulation of molecular energies,}
  {\protect\JournalTitle{Physical Review X}} \textbf{6}, 031007 (2016).

\bibitem{kandala2017hardware}
A.~Kandala, A.~Mezzacapo, K.~Temme, M.~Takita, M.~Brink, J.~M. Chow, and J.~M.
  Gambetta, \enquote{Hardware-efficient variational quantum eigensolver for
  small molecules and quantum magnets,} {\protect\JournalTitle{Nature}}
  \textbf{549}, 242--246 (2017).

\bibitem{yung2014trappedion}
M.-H. Yung, J.~Casanova, A.~Mezzacapo, J.~Mcclean, L.~Lamata, A.~Aspuru-Guzik,
  and E.~Solano, \enquote{From transistor to trapped-ion computers for quantum
  chemistry,} {\protect\JournalTitle{Sci. Rep.}} \textbf{4}, 1--7 (2014).

\bibitem{shen2017quantum}
Y.~Shen, X.~Zhang, S.~Zhang, J.-N. Zhang, M.-H. Yung, and K.~Kim,
  \enquote{Quantum implementation of the unitary coupled cluster for simulating
  molecular electronic structure,} {\protect\JournalTitle{Physical Review A}}
  \textbf{95}, 020501 (2017).

\bibitem{kokail2019self}
C.~Kokail, C.~Maier, R.~van Bijnen, T.~Brydges, M.~K. Joshi, P.~Jurcevic, C.~A.
  Muschik, P.~Silvi, R.~Blatt, C.~F. Roos \emph{et~al.},
  \enquote{Self-verifying variational quantum simulation of lattice models,}
  {\protect\JournalTitle{Nature}} \textbf{569}, 355--360 (2019).

\bibitem{biamonte2017quantum}
J.~Biamonte, P.~Wittek, N.~Pancotti, P.~Rebentrost, N.~Wiebe, and S.~Lloyd,
  \enquote{Quantum machine learning,} {\protect\JournalTitle{Nature}}
  \textbf{549}, 195--202 (2017).

\bibitem{lubasch2020variational}
M.~Lubasch, J.~Joo, P.~Moinier, M.~Kiffner, and D.~Jaksch, \enquote{Variational
  quantum algorithms for nonlinear problems,} {\protect\JournalTitle{Physical
  Review A}} \textbf{101}, 010301 (2020).

\bibitem{liang2020variational}
J.-M. Liang, S.-Q. Shen, M.~Li, and L.~Li, \enquote{Variational quantum
  algorithms for dimensionality reduction and classification,}
  {\protect\JournalTitle{Physical Review A}} \textbf{101}, 032323 (2020).

\bibitem{benedetti2020hardware}
M.~Benedetti, M.~Fiorentini, and M.~Lubasch, \enquote{Hardware-efficient
  variational quantum algorithms for time evolution,}
  {\protect\JournalTitle{Physical Review Research}} \textbf{3}, 033083 (2021).

\bibitem{barron2020measurement}
G.~S. Barron and C.~J. Wood, \enquote{Measurement error mitigation for
  variational quantum algorithms,} {\protect\JournalTitle{arXiv preprint
  arXiv:2010.08520}}  (2020).

\bibitem{su2020error}
D.~Su, R.~Israel, K.~Sharma, H.~Qi, I.~Dhand, and K.~Br{\'a}dler,
  \enquote{Error mitigation on a near-term quantum photonic device,}
  {\protect\JournalTitle{Quantum}} \textbf{5}, 452 (2021).

\bibitem{knill2001scheme}
E.~Knill, R.~Laflamme, and G.~J. Milburn, \enquote{A scheme for efficient
  quantum computation with linear optics,} {\protect\JournalTitle{Nature}}
  \textbf{409}, 46--52 (2001).

\bibitem{hong2020experimental}
S.~Hong, C.~H. Park, Y.-H. Choi, Y.-S. Kim, Y.-W. Cho, K.~Oh, and H.-T. Lim,
  \enquote{Experimental implementation of arbitrary entangled operations,}
  {\protect\JournalTitle{New Journal of Physics}} \textbf{22}, 093070 (2020).

\bibitem{wang2019boson}
H.~Wang, J.~Qin, X.~Ding, M.-C. Chen, S.~Chen, X.~You, Y.-M. He, X.~Jiang,
  L.~You, Z.~Wang \emph{et~al.}, \enquote{Boson sampling with 20 input photons
  and a 60-mode interferometer in a $10^{14}$-dimensional {H}ilbert space,}
  {\protect\JournalTitle{Physical Review Letters}} \textbf{123}, 250503 (2019).

\bibitem{barreiro2008beating}
J.~T. Barreiro, T.-C. Wei, and P.~G. Kwiat, \enquote{Beating the channel
  capacity limit for linear photonic superdense coding,}
  {\protect\JournalTitle{Nature Physics}} \textbf{4}, 282--286 (2008).

\bibitem{yoo2018experimental}
J.~Yoo, Y.~Choi, Y.-W. Cho, S.-W. Han, S.-Y. Lee, S.~Moon, K.~Oh, and Y.-S.
  Kim, \enquote{Experimental preparation and characterization of
  four-dimensional quantum states using polarization and time-bin modes of a
  single photon,} {\protect\JournalTitle{Optics Communications}} \textbf{419},
  30--35 (2018).

\bibitem{wang201818}
X.-L. Wang, Y.-H. Luo, H.-L. Huang, M.-C. Chen, Z.-E. Su, C.~Liu, C.~Chen,
  W.~Li, Y.-Q. Fang, X.~Jiang \emph{et~al.}, \enquote{18-qubit entanglement
  with six photon's three degrees of freedom,} {\protect\JournalTitle{Physical
  Review Letters}} \textbf{120}, 260502 (2018).

\bibitem{NC:11:QCQI}
M.~A. Nielsen and I.~L. Chuang, \emph{Quantum Computation and Quantum
  Information: 10th Anniversary Edition} (Cambridge University Press, USA,
  2011), 10th ed.

\bibitem{Hoe:63:JASA}
W.~Hoeffding, \enquote{Probability inequalities for sums of bounded random
  variables,} {\protect\JournalTitle{J. Am. Stat. Assoc.}} \textbf{58}, 13--30
  (1963).

\bibitem{SWM:20:Qua}
R.~Sweke, F.~Wilde, J.~Meyer, M.~Schuld, P.~K. Faehrmann, B.~Meynard-Piganeau,
  and J.~Eisert, \enquote{Stochastic gradient descent for hybrid
  quantum-classical optimization,} {\protect\JournalTitle{Quantum}} \textbf{4},
  314 (2020).

\bibitem{SIK:20:Qua}
J.~Stokes, J.~Izaac, N.~Killoran, and G.~Carleo, \enquote{Quantum natural
  gradient,} {\protect\JournalTitle{Quantum}} \textbf{4}, 269 (2020).

\bibitem{UB:20:arx}
A.~Uvarov and J.~Biamonte, \enquote{On barren plateaus and cost function
  locality in variational quantum algorithms,} {\protect\JournalTitle{Journal
  of Physics A: Mathematical and Theoretical}} \textbf{54}, 245301 (2021).

\bibitem{WFC:21:arx}
S.~Wang, E.~Fontana, M.~Cerezo, K.~Sharma, A.~Sone, L.~Cincio, and P.~J. Coles,
  \enquote{Noise-induced barren plateaus in variational quantum algorithms,}
  {\protect\JournalTitle{arXiv:2007.14384}}  (2020).

\bibitem{RBM:18:arx}
J.~Romero, R.~Babbush, J.~R. McClean, C.~Hempel, P.~Love, and A.~Aspuru-Guzik,
  \enquote{Strategies for quantum computing molecular energies using the
  unitary coupled cluster ansatz,} {\protect\JournalTitle{Quantum Science and
  Technology}} \textbf{4}, 014008 (2018).

\bibitem{PSP:21:QIP}
A.~Pellow-Jarman, I.~Sinayskiy, A.~Pillay, and F.~Petruccione, \enquote{A
  comparison of various classical optimizers for a variational quantum linear
  solver,} {\protect\JournalTitle{Quantum Inf. Process.}} \textbf{20}, 202
  (2021).

\bibitem{CSV:21:npjQI}
M.~Cerezo, A.~Sone, T.~Volkoff, L.~Cincio, and P.~J. Coles, \enquote{Cost
  function dependent barren plateaus in shallow parametrized quantum circuits,}
  {\protect\JournalTitle{Nature Commun.}} \textbf{12}, 1791 (2021).

\bibitem{LCS:21:arx}
M.~Larocca, P.~Czarnik, K.~Sharma, G.~Muraleedharan, P.~J. Coles, and
  M.~Cerezo, \enquote{Diagnosing barren plateaus with tools from quantum
  optimal control,} {\protect\JournalTitle{arXiv: 2105.14377}}  (2021).

\bibitem{MZO:20:qua}
F.~B. Maciejewski, Z.~Zimbor{\'{a}}s, and M.~Oszmaniec, \enquote{Mitigation of
  readout noise in near-term quantum devices by classical post-processing based
  on detector tomography,} {\protect\JournalTitle{Quantum}} \textbf{4}, 257
  (2020).

\bibitem{rehman2021entanglementfree}
J.~ur~Rehman and H.~Shin, \enquote{Entanglement-free parameter estimation of
  generalized {P}auli channels,} {\protect\JournalTitle{Quantum}} \textbf{5},
  490 (2021).

\bibitem{RJK:18:SR}
J.~ur~Rehman, Y.~Jeong, J.~S. Kim, and H.~Shin, \enquote{Holevo capacity of
  discrete {W}eyl channels,} {\protect\JournalTitle{Sci. Rep.}} \textbf{8},
  17457 (2018).

\bibitem{RJS:19:PRA}
J.~ur~Rehman, Y.~Jeong, and H.~Shin, \enquote{Directly estimating the {H}olevo
  capacity of discrete {W}eyl channels,} {\protect\JournalTitle{Physical Review
  A}} \textbf{99}, 042312 (2019).

\bibitem{SIU:20:JPAMT}
K.~Siudzi{\'{n}}ska, \enquote{Classical capacity of generalized {P}auli
  channels,} {\protect\JournalTitle{J. Phys. A Math. Theor.}} \textbf{53},
  445301 (2020).

\bibitem{wang2016experimental}
X.-L. Wang, L.-K. Chen, W.~Li, H.-L. Huang, C.~Liu, C.~Chen, Y.-H. Luo, Z.-E.
  Su, D.~Wu, Z.-D. Li \emph{et~al.}, \enquote{Experimental ten-photon
  entanglement,} {\protect\JournalTitle{Physical Review Letters}} \textbf{117},
  210502 (2016).

\bibitem{pramanik2019revealing}
T.~Pramanik, Y.-W. Cho, S.-W. Han, S.-Y. Lee, Y.-S. Kim, and S.~Moon,
  \enquote{Revealing hidden quantum steerability using local filtering
  operations,} {\protect\JournalTitle{Physical Review A}} \textbf{99}, 030101
  (2019).

\bibitem{pramanik2019nonlocal}
T.~Pramanik, Y.-W. Cho, S.-W. Han, S.-Y. Lee, S.~Moon, and Y.-S. Kim,
  \enquote{Nonlocal quantum correlations under amplitude damping decoherence,}
  {\protect\JournalTitle{Physical Review A}} \textbf{100}, 042311 (2019).

\bibitem{yoon2019experimental}
J.~Yoon, T.~Pramanik, B.-K. Park, Y.-W. Cho, S.-Y. Lee, S.~Kim, S.-W. Han,
  S.~Moon, and Y.-S. Kim, \enquote{Experimental comparison of various quantum
  key distribution protocols under reference frame rotation and fluctuation,}
  {\protect\JournalTitle{Optics Communications}} \textbf{441}, 64--68 (2019).

\bibitem{park2015high}
B.~K. Park, Y.-S. Kim, O.~Kwon, S.-W. Han, and S.~Moon,
  \enquote{High-performance reconfigurable coincidence counting unit based on a
  field programmable gate array,} {\protect\JournalTitle{Applied Optics}}
  \textbf{54}, 4727--4731 (2015).

\bibitem{park2021arbitrary}
B.~K. Park, Y.-S. Kim, Y.-W. Cho, S.~Moon, and S.-W. Han, \enquote{Arbitrary
  configurable 20-channel coincidence counting unit for multi-qubit quantum
  experiment,} {\protect\JournalTitle{Electronics}} \textbf{10}, 569 (2021).

\bibitem{hamamura2020efficient}
I.~Hamamura and T.~Imamichi, \enquote{Efficient evaluation of quantum
  observables using entangled measurements,} {\protect\JournalTitle{npj Quantum
  Information}} \textbf{6}, 1--8 (2020).

\end{thebibliography}
\end{document}



\section{Hamiltonian of He--H$^+$ with respect to the interatomic distance}


According to the second-order quantization representation, any Hamiltonian can be represented as
\begin{eqnarray}
	H = \sum_{p,q} h_{p,q} a^\dag_{p}a_q + \sum_{p,q,r,s} h_{p,q,r,s} a^\dag_{p}a^\dag_q a_r a_s + \cdots,
\label{eq_aa}
\end{eqnarray}
where $a^\dag$ and $a$ represent the creation and annihilation operators, respectively. By Jordan-Wigner transformation, \eqref{eq_aa} an be presented as a linear summations of Pauli operators as
\begin{eqnarray}
	H = \sum_{j} w_{j} \cdot\M{\sigma}_{j} + \sum_{jk} w_{jk} \cdot\M{\sigma}_{j}\otimes\M{\sigma}_{k} + \cdots.
\label{eq_ham}
\end{eqnarray}
The coefficients $w_j, w_{jk}, \cdots$ are calculated during the transformation.

Here, we have calculated the molecular Hamiltonians with respect to the interatomic distance for He--H$^+$ using Psi4 module, and listed the coefficients in Table~\ref{table_Hamitonian}~\cite{turney2012psi4}. $X,Y,Z,I$ here stand for the Pauli matrices $\sigma_{x}, \sigma_{y}, \sigma_{z}$ and the identity operators, respectively. Note that, while the Hamiltonian has nine Pauli string terms, it only requires four measurement settings (Group 1$\sim$4) to obtain all the values.

\begin{table}[h!]
\setlength{\tabcolsep}{0.05in}
\centering
\begin{tabular}{|c|c|c|c|c|c|c|c|c|c|}
\hline
\multicolumn{1}{|c|}{} & \multicolumn{4}{c|}{Group 1}& \multicolumn{2}{c|}{Group 2}& \multicolumn{2}{c|}{Group 3}& \multicolumn{1}{c|}{~~Group 4~~}\\
\hline
~~R~[\AA]~~ & II & IZ & ZI & ZZ & IX & ZX & XI & XZ & XX \\
\hline
0.05 & 33.9557 & -2.4784 & -2.4784 & 0.2746 & -0.1515 & 0.1515 & -0.1515 & 0.1515 & 0.1412 \\
0.1 & 13.3605 & -2.4368 & -2.4368 & 0.2081 & -0.1626 & 0.1626 & -0.1626 & 0.1626 & 0.2097 \\
0.2 & 3.633 & -2.2899 & -2.2899 & 0.1176 & -0.1405 & 0.1405 & -0.1405 & 0.1405 & 0.3027 \\
0.5 & -2.3275 & -1.5236 & -1.5236 & 0.1115 & -0.157 & 0.157 & -0.157 & 0.157 & 0.3309 \\
0.7 & -3.3893 & -1.2073 & -1.2073 & 0.1626 & -0.1968 & 0.1968 & -0.1968 & 0.1968 & 0.3052 \\
0.9 & -3.8505 & -1.0466 & -1.0466 & 0.2356 & -0.2288 & 0.2288 & -0.2288 & 0.2288 & 0.2613 \\
1.1 & -4.0539 & -0.982 & -0.982 & 0.3225 & -0.243 & 0.243 & -0.243 & 0.243 & 0.2053 \\
1.5 & -4.1594 & -0.991 & -0.991 & 0.4945 & -0.2086 & 0.2086 & -0.2086 & 0.2086 & 0.0948 \\
2 & -4.1347 & -1.0605 & -1.0605 & 0.6342 & -0.1119 & 0.1119 & -0.1119 & 0.1119 & 0.0212 \\
2.5 & -4.0918 & -1.1128 & -1.1128 & 0.701 & -0.0454 & 0.0454 & -0.0454 & 0.0454 & 0.0032 \\
\hline
\hline
\end{tabular}
\caption{The table of Pauli operators and weights constituting Hamiltonian in respect to interatomic distance~\cite{peruzzo2014photonic}. Four measurement settings (Group 1$\sim$4) are required for obtain all the nine Pauli strings.}
\label{table_Hamitonian}
\end{table}

\section{The test results of various classical optimizers}



\begin{figure}[h!]
\includegraphics[angle=0,width=5in]{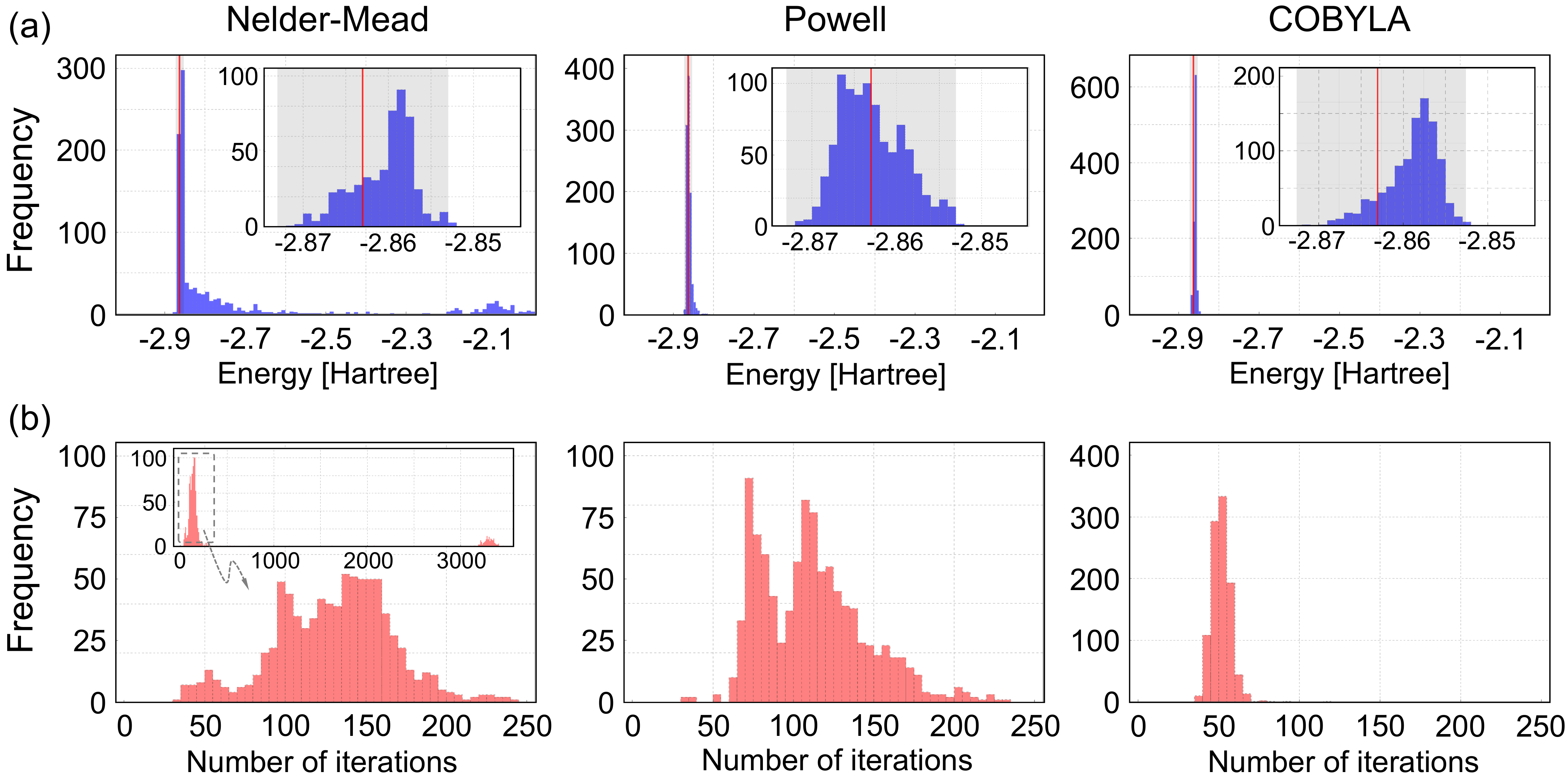}
\caption{The classical computer emulation test results of the classical optimizers with 1,000 trials. (a) the estimated energy and (b) the number of iterations. In (a) the red solid line represents theoretical value of our simulated Hamiltonian, R=0.9\AA. The gray region shows the range of the objective value tolerance condition, ftol=0.01, which is corresponding to success criteria.}
\label{vqe_histogram_sim}
\end{figure}

We have evaluated three classical optimizers with 1,000 trials of classical computer emulation. The initial parameters have set randomly for each emulation and experiments. Figure~\ref{vqe_histogram_sim} shows the histogram of each classical optimizer in terms of (a) the estimated energy, and (b) the number of iterations. In Fig.~\ref{vqe_histogram_sim} (a), the red line indicate the theory value with the emulation condition of interatomic distance $R=0.9$\AA. The gray region represents the area of success criteria corresponding to tolerance of optimizer, ftol = 0.01. The mean values of the estimated energy $\langle H \rangle$ with Nelder-Mead, Powell and COBYLA methods are -2.861~$\pm$~0.004, -2.863~$\pm$~0.004 and -2.860~$\pm$~0.003. The differences from the theoretical value are negligible. From Fig. \ref{vqe_histogram_sim} (b), we can notice that the mean iteration numbers of VQE with Nelder-Mead, Powell and COBYLA method are 130, 112, and 51, respectively. In the iteration number histogram of Nelder-Mead method, inner histogram has some cases over 3,000 iterations due to trap in local minimums. We note that each iteration requires one QPU call, which takes up most time in our VQE loop, so that we select COBYLA method as our classical optimizer. 

\begin{figure}[h!]
\includegraphics[angle=0,width=5in]{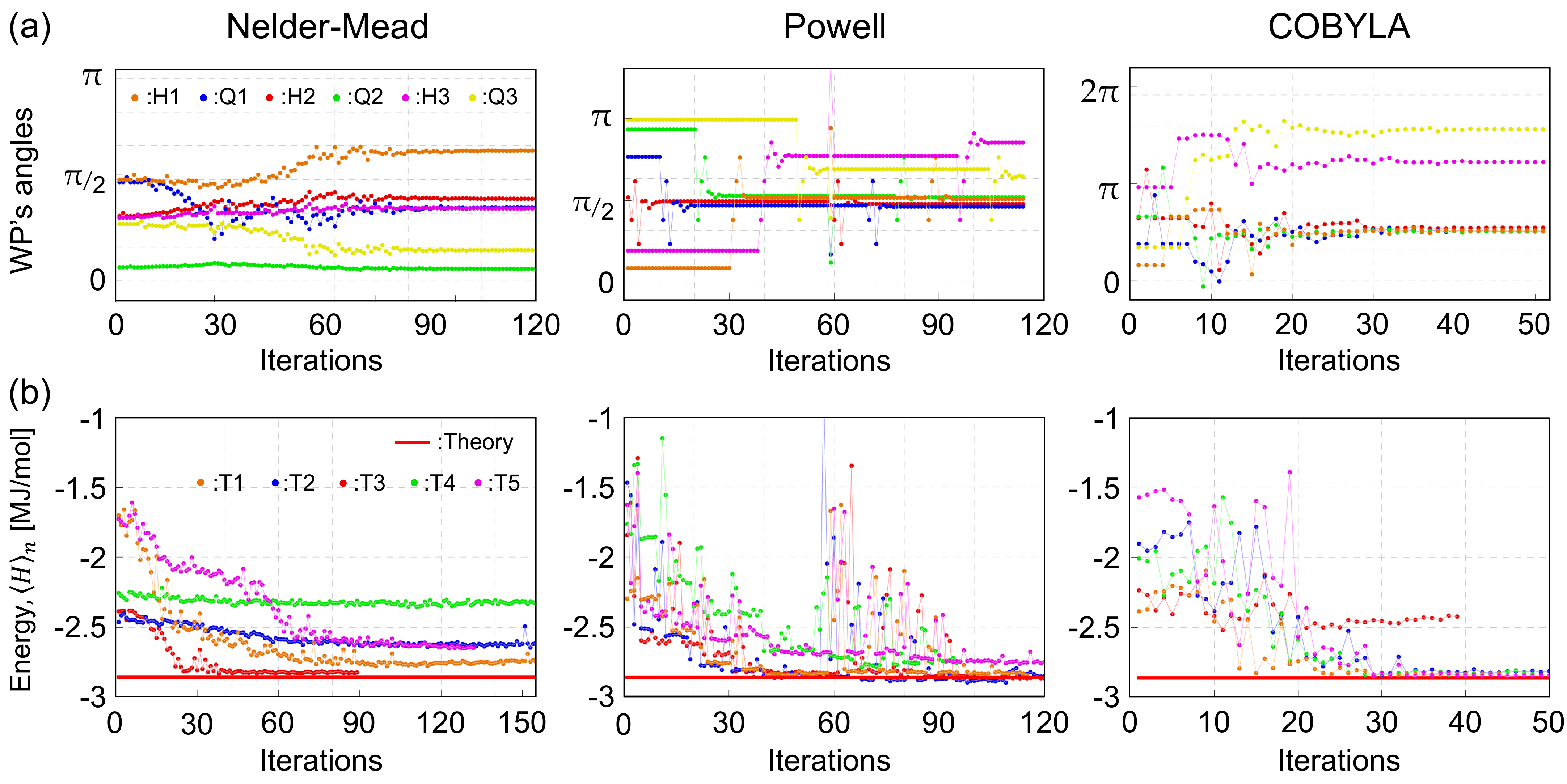}
\caption{The experimental results of real VQE runs with different classical optimizers. (a) the waveplate (WP) angles change during the single VQE run. (b) the estimated energy changes during the iteration in five independent VQE runnings.}
\label{vqe_histogram_exp}
\end{figure}

Figure~\ref{vqe_histogram_exp} shows the experimental results of each classical optimizers. Figure~\ref{vqe_histogram_exp}(a) shows the change of input parameters (WP's angles) during a single VQE run. Figure~\ref{vqe_histogram_exp}(b) shows that the estimated bound energy at $R~=~0.9$\AA~ with respect to the number of iterations in five independent VQE runs. For five independent trials, Nelder-Mead, Powell, and COBYLA have successfully found the ground state energy 1, 4, and 4 times, respectively.
$\\$

$\\$

\newpage

\section{Photonic four dimensional quantum state generation and measurement}

\begin{figure}[h!]
\includegraphics[angle=0,width=5.2in]{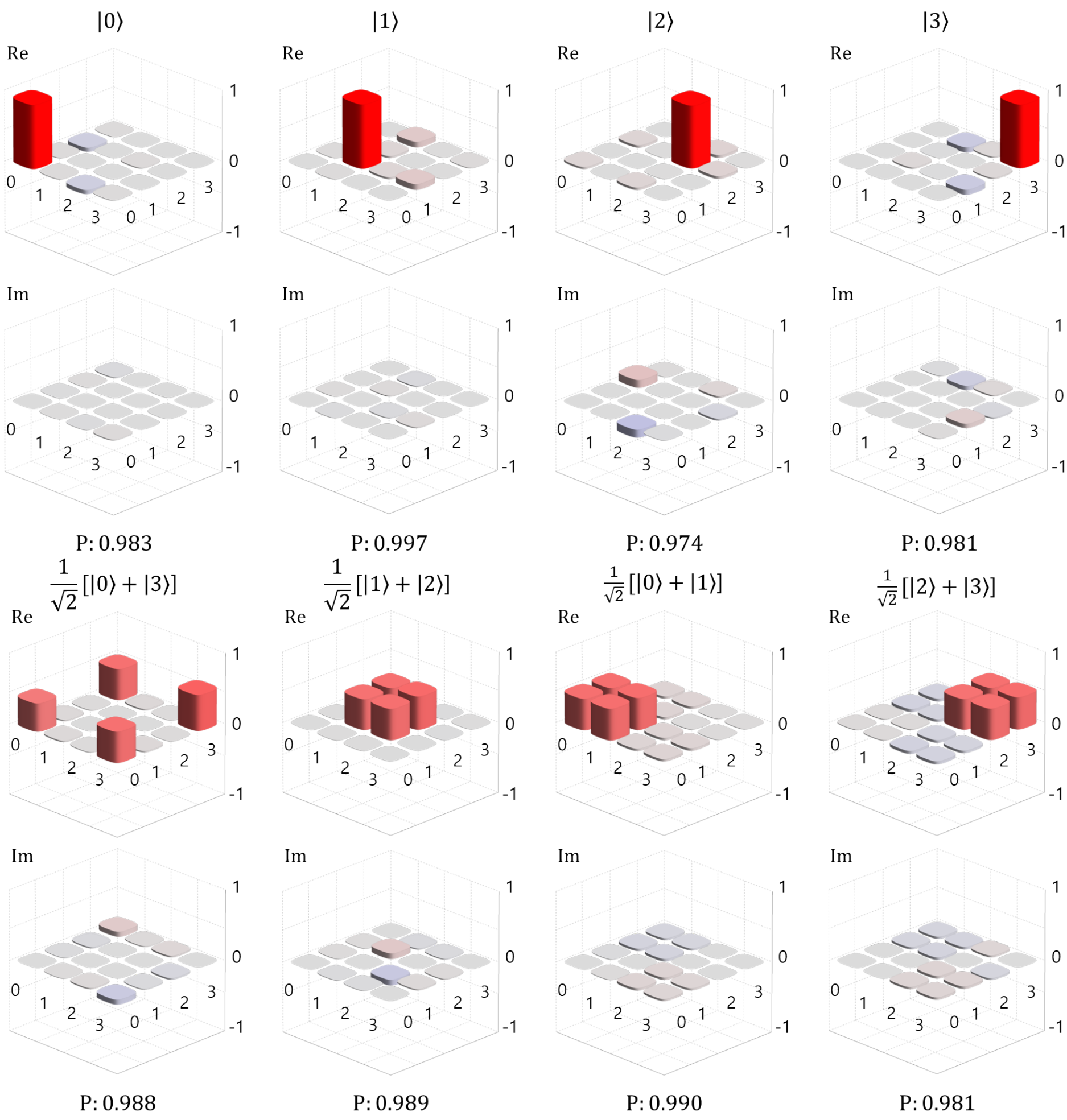}
\caption{The QST and its state purity results of given ququart states (part 1).}
\label{Ququarts_qst_1}
\end{figure}

\begin{figure}[h!]
\includegraphics[angle=0,width=5.2in]{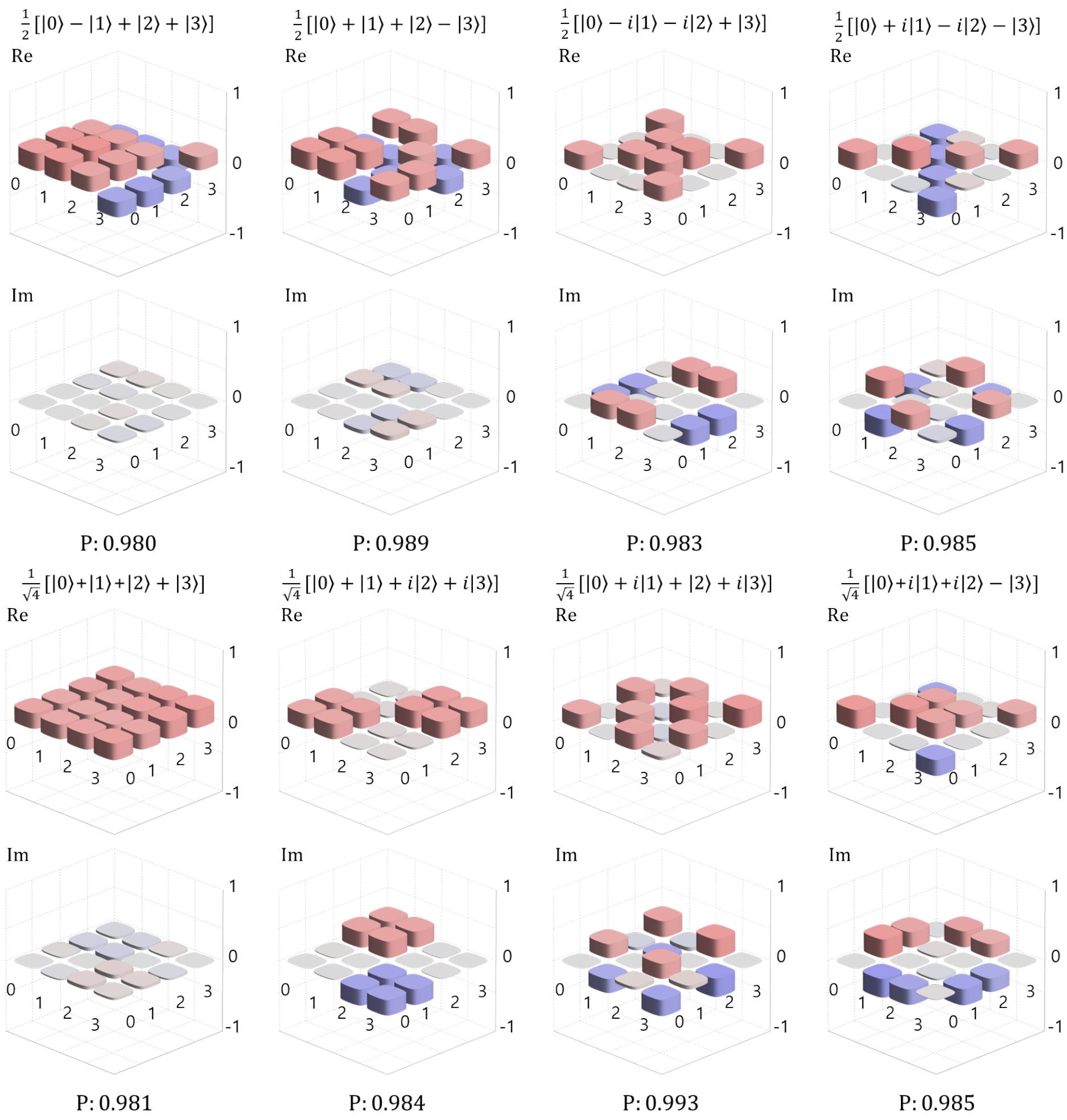}
\caption{The QST and its state purity results of given ququart states (part 2).}
\label{Ququarts_qst_2}
\end{figure}

We have tested our QPU by generating and measuring various four-dimensional quantum states. Here,  we have generated the ququart states in mutually unbiased bases and checked the state purity through quantum state tomography (QST) \cite{yoo2018experimental}. Figure \ref{Ququarts_qst_1} and \ref{Ququarts_qst_2} show the QST results with the values of state purity. For clear description, we present $\vert aH \rangle, \vert aV \rangle, \vert bH \rangle $ and $\vert bV \rangle$ to $\vert 0 \rangle, \vert 1 \rangle, \vert 2 \rangle $and $\vert 3 \rangle$, respectively.

\newpage

$\\\\$

